\documentclass{ws-mpla}
\usepackage{amssymb}
\usepackage{revsymb}
\usepackage{epsfig}
\usepackage{graphicx}
\usepackage{lscape}
\usepackage[square,comma,numbers]{natbib}
\begin{document}

\markboth{Nissim Kanekar}
%{Probing fundamental constant evolution with radio spectroscopy}
{Do the fundamental constants change with time ?}

%%%%%%%%%%%%%%%%%%%%% Publisher's Area please ignore %%%%%%%%%%%%%%
\catchline{}{}{}{}{}
%%%%%%%%%%%%%%%%%%%%%%%%%%%%%%%%%%%%%%%%%%%%%%%%%%%%%%%%%%%%%%%%%%%

\setcounter{equation}{0}
\setcounter{figure}{0}
\setcounter{footnote}{0}
\setcounter{section}{0}
\setcounter{table}{0}

%\citestyle{aa}

% Some definitions I use in these instructions.

\newcommand{\dal}{\ensuremath{\left[\Delta \alpha/ \alpha \right]}}
\newcommand{\dmu}{\ensuremath{\left[\Delta \mu/\mu \right]}}
\newcommand{\hi}{H{\sc i}}
\newcommand{\htwo}{${\rm H}_2$}

\title{DO THE FUNDAMENTAL CONSTANTS CHANGE WITH TIME ?}
\author{\footnotesize NISSIM KANEKAR}
\address{National Radio Astronomy Observatory, 1003 Lopezville Road, 
Socorro, NM 87801, USA\\ nkanekar@nrao.edu}
\maketitle
\pub{}{}

\begin{abstract}
Comparisons between the redshifts of spectral lines from cosmologically-distant 
galaxies can be used to probe temporal changes in low-energy fundamental constants like
the fine structure constant and the proton-electron mass ratio. In this article, I 
review the results from, and the advantages and disadvantages of, the best techniques 
using this approach, before focussing on a new method, based on conjugate satellite OH lines, 
that appears to be less affected by systematic effects and hence holds much promise for the 
future.
\end{abstract}

\ccode{PACS Nos.: 98.80.Es,06.20.Jr,33.20.Bx,98.58.-w}

\section{Introduction}
\label{sec:intro}

In modern higher-dimensional extensions of the standard model of particle physics,
low-energy fundamental constants like the fine structure constant $\alpha$, the 
proton-electron mass ratio $\mu \equiv m_p/m_e$, etc, are expected to be dynamical quantities
that show spatio-temporal evolution (e.g.  \cite{marciano84,damour94}). The detection of such 
changes provides an avenue to explore new physics, especially important because most 
other predictions of these models lie at inaccessibly high energies ($\gtrsim 10^{19}$~GeV). 
However, the timescales of these putative changes are unknown, implying that it is 
necessary to search for evolution over as wide a range of timescales as possible (especially
as it is not even clear that any evolution will be monotonic). It is also important to 
probe changes in different constants: fractional changes in $\mu$ are expected to be far 
larger than those in $\alpha$, by factors of $\sim 50- 500$ (e.g. \cite{calmet02,langacker02}). 

On short timescales (a few years), laboratory experiments, usually based on frequency 
comparisons between atomic clocks and other optical frequency standards have provided the 
strongest constraints on changes in (especially) the fine structure constant. For example,
a recent comparison between $^{199}$Hg$^+$ and $^{27}$Al$^+$ clocks found 
$\dot{\alpha}/\alpha < 4.6 \times 10^{-17}$~yr$^{-1}$ over one year \cite{rosenband08}. 
In the case of the proton-electron mass ratio, \cite{shelkolnikov08} obtained 
$\dot{\mu}/\mu < 1.2 \times 10^{-13}$~yr$^{-1}$, comparing frequencies derived from 
molecular and cesium clocks over a two-year period. Note that all limits quoted 
in this paper are at $2\sigma$ significance.

Besides the impressive precision of these (and other) laboratory results, 
it should be noted that they allow direct control over systematic effects. Unfortunately, 
they are limited to probing changes in the constants over timescales of only a few years. 
Various geological methods have allowed the extension of such studies to lookback times 
of a few billion years; most prominent amongst these are measurements of relative isotopic 
abundances in the Oklo natural fission reactor and studies of the time variation of the 
$\beta$-decay rate (e.g. \cite{uzan03}). However, while sensitive constraints on changes 
in $\alpha$ (over $\sim 1.8-4$~Gyr) have been obtained from both these methods, they 
typically are model-dependent and involve assumptions about the constancy of other 
parameters (e.g. \cite{fujii03,gould06,petrov06}). For example, \cite{gould06} obtain the 
$2\sigma$ limit $-0.24 \times 10^{-7} < \dal < 0.11 \times 10^{-7}$ from 
a re-analysis of Oklo nuclear yields, but emphasize that this depends critically on 
the assumption that $\alpha$ alone changes with time; any changes in the nuclear 
potential could significantly affect the result.

Astrophysical techniques allow tests of changes in the fundamental constants over 
large lookback times, out to significant fractions of the age of the Universe. 
The methods can be broadly divided into the following three categories: 
(1)~measurements of the abundances of elements formed during Big Bang nucleosynthesis (BBN)
\cite{kolb86}, (2)~observations of anisotropies in the cosmic microwave background 
\cite{hannestad99,kaplinghat99} and (3)~comparisons between the redshifts of 
spectral lines from distant galaxies \cite{savedoff56}. The first two of these
probe the largest lookback times, but the results are not very sensitive due to 
degeneracies with the cosmological parameters, which are not {\it a priori} known 
(e.g. \cite{rocha04}). Additionally, most BBN studies are 
highly model-dependent as the dependence of the nuclear binding energies on the 
fundamental constants is still not known \cite{dent07}. In this article, I will 
concentrate on the third category, the use of astronomical spectroscopy to probe 
changes in three constants, $\alpha$, $\mu \equiv m_p/m_e$ and the proton g-factor 
ratio $g_p$ (or combinations of these quantities). After discussing 
the results from, and advantages and disadvantages of, the main techniques currently 
used in this endeavour, I will focus on a new technique, based on ``conjugate'' 
satellite hydroxyl (OH) lines, that appears to be less affected by systematic effects. 
Finally, I will discuss prospects for such studies of fundamental constant evolution,
based on new telescopes and instruments over the next decade.

\section{Redshifted spectral lines: background}
\label{sec:astro-spec}

The idea that comparisons between the velocities of different transitions arising 
in a distant galaxy might be used to test for spatio-temporal dependences 
in the fundamental constants was first suggested more than half a century ago 
\cite{savedoff56} and is conceptually quite simple. Different spectral transitions 
arise from different physical mechanisms (e.g. interactions of fine structure, hyperfine 
structure, rotational or other types) and the line frequencies have different dependences 
on the fundamental constants. For example, hyperfine line frequencies 
are proportional to $g_p \mu \alpha^2$, while rotational line frequencies are 
proportional to $\mu$. If the fundamental constants are different at different 
space-time locations (e.g. in external galaxies), the rest frequency of a transition 
would be different at these locations from the value measured in terrestrial 
laboratories. Different spectral transitions arising in a given gas cloud 
(or galaxy) would then yield different systemic velocities, as the incorrect
(laboratory) rest frequency would be used for each line to estimate the cloud
velocity. A comparison between the line velocities and the known dependence of
the line rest frequency on the constants can then be used to determine the 
fractional change in a given constant between the space-time locations of the cloud
and the Earth. For example, a comparison between hyperfine and rotational velocities 
is sensitive to changes in $g_p \alpha^2$, and one between hyperfine and fine structure
transitions, to changes in $g_p \mu \alpha^2$. Most such studies are carried out 
in absorption (or stimulated emission), against distant quasars, because absorption
lines can be detected out to high redshifts with undiminished intensity and are also 
usually much narrower than emission lines, allowing precision measurements of the 
line redshift (but see \cite{bahcall04}).

A critical assumption here is that an observed velocity offset between different 
transitions can be ascribed to changes in the fundamental constants, i.e. that 
``local'' velocity offsets (e.g. due to motions within galaxies or clouds) are 
negligible. This is manifestly incorrect as transitions from different species 
often arise from different parts of a gas cloud, at different velocities.
Even transitions from one species need not arise at the same velocity, if 
the cloud is not in equilibrium.  Intra-cloud velocities of $\sim 10$~km/s are 
common in astrophysical circumstances; these would yield systematic effects 
of order $(v/c) \sim few \times 10^{-5}$ on estimates of fractional changes in 
the constants (e.g. \cite{wiklind97,carilli00}). As will be seen later, the best 
present estimates of fractional changes in $\alpha$ and $\mu$ are of this order, 
implying that such local offsets cannot be neglected. Two approaches have been 
used to handle such systematics: (1)~a statistically-large sample of spectral lines, 
to average out local effects (e.g. the many-multiplet method; \cite{dzuba99}) and 
(2)~the use of certain spectral lines, where the physics of the line mechanism 
causes such local effects to be negligible (e.g. the conjugate satellite OH approach;
\cite{kanekar04b,darling04}). 

\section{Optical techniques}
\label{sec:optical}

Most studies of fundamental constant evolution based on redshifted spectral lines 
have used optical techniques, partly, no doubt, due to the profusion of strong 
ultraviolet transitions that are redshifted into optical wavebands. The three principal 
approaches are:

\subsection{Alkali doublet method}
\label{sec:optical-doublets}

For three decades, redshifted alkali doublet (AD) lines provided the main spectroscopic 
probe of the long-term evolution of the fine structure constant (e.g. 
\cite{bahcall67,ivanchik99}).  The fractional separation between the doublet 
wavelengths is proportional to $\alpha^2$ and the measured line redshifts 
thus directly provide an estimate of $\dal$ \citep{savedoff56}.  The best 
present result from this technique is $\dal < 2.6 \times 10^{-5}$, based on High Resolution 
Echelle Spectrograph (HIRES) absorption spectra of 21~Si{\sc iv} doublets at $2 < z < 3$, 
with the Keck telescope \cite{murphy01c}. While \cite{chand05} obtained 
$\dal < 0.86 \times 10^{-5}$ from Very Large Telescope Ultraviolet Echelle Spectrograph 
(VLT-UVES) spectra of 15~Si{\sc iv} doublets at $1.6 < z < 2.9$, it has been pointed out 
that their $\chi^2$ curves contain large fluctuations, making the result unreliable 
\cite{murphy08b}. Although the sensitivity of the AD method is nearly an order 
of magnitude poorer than that of the many-multiplet method (discussed next), the fact
that the two doublet lines have the same shape provides a useful test for systematic 
effects. Further, the doublet lines lie at nearby wavelengths, and hence usually 
on the same echelle order, implying that relative wavelength calibration across
orders is not an issue. \cite{murphy01d} note that differential isotopic saturation of 
the Si{\sc iv} lines must be taken into account when using saturated lines in this method; 
even better would be to restrict the analysis to unsaturated Si{\sc iv} absorbers alone.

\subsection{The many-multiplet method}
\label{sec:mm}

The many-multiplet (MM) method, devised by \cite{dzuba99}, is based on the fact that 
relativistic first-order corrections imply that the wavelengths of fine structure 
transitions in different species have different dependences on $\alpha$. Thus, unlike the 
AD method, the MM method is based on comparisons between transitions 
from {\it different} species and hence must be carried out on large absorber samples 
to average out local effects. This has resulted in a significant improvement 
in sensitivity to fractional changes in $\dal$ over the alkali doublet approach, 
albeit perhaps with a concomitant increase in systematic effects (see, e.g., 
\cite{bahcall04,murphy01d,murphy03,molaro07}). This is also, at present, the 
only technique that finds evidence for a change in the fine structure constant:
\cite{murphy04} used Keck-HIRES data to obtain $\dal = \left[-0.573 \pm 0.113\right] 
\times 10^{-5}$ from 143~absorbers in the redshift range $0.2 < z < 4.2$, implying that 
$\alpha$ was smaller at earlier times (see also \cite{murphy03}). A slew of later 
VLT-UVES observations have claimed strong constraints on $\dal$, from either the MM 
method or its variants (e.g. the SIDAM method; \cite{quast04,levshakov07}), some of which are 
inconsistent with the Keck-HIRES result (e.g. \cite{chand04}). For example, \cite{molaro07,levshakov07} 
use the SIDAM method to obtain $\dal = (5.7 \pm 2.7) \times 10^{-6}$ and 
$\dal = (-0.12 \pm 1.79) \times 10^{-6}$ from absorbers at $z \sim 1.84$ and $z \sim 1.15$, 
respectively. However, \cite{murphy08b} point out that the $\chi^2$ curves of \cite{chand04} 
show large fluctuations, causing them to under-estimate their errors by a factor of 
$\gtrsim 3$. \cite{murphy08b} also argue that some of the other VLT-UVES results are likely 
to be unreliable, notably due to the fitting of insufficient components \cite{quast04,levshakov07}. 
Conversely, \cite{molaro07} find non-gaussian tails in the distribution of the Keck-HIRES 
data of \cite{murphy03,murphy04}, as well as correlations in the data, suggesting that they 
too may have under-estimated their errors.

Excellent discussions of the systematic effects inherent in the MM (and AD) methods 
are provided by \cite{murphy01d,murphy03}, who find no evidence that their Keck-HIRES 
result may be affected by such systematics (but see \cite{molaro07}). Isotopic 
abundance variation with redshift is perhaps the most important source of 
systematic effects in the MM method, as lines from multiple isotopes are 
blended for most species (especially Mg), and terrestrial isotopic abundances are assumed 
in order to determine the central line wavelength \cite{murphy04,ashenfelter04}. For 
example, higher $^{25,26}$Mg fractional abundances in the $z < 1.8$ sample of \cite{murphy04} 
could yield the observed negative $\dal$ for this sample. Interestingly enough, \cite{molaro07} 
find that the results of \cite{murphy03,murphy04} appear to be dominated by the sample that uses 
Mg{\sc ii} lines: the sub-sample of \cite{murphy03} that uses the Mg{\sc ii} lines yields 
a $\beta$-trimmed mean of $\dal = (-0.48 \pm 0.12) \times 10^{-5}$, while the sub-sample 
without these lines gives $\dal = (-0.11 \pm 0.17) \times 10^{-5}$ \cite{molaro07}.
Further, the isotopic structure is only known for a few of the transitions (e.g. Mg, Si) 
used in the full analysis \cite{murphy01c}. Unfortunately, it will be difficult to resolve this 
issue by direct observations, while indirect arguments depend on the details of galactic 
chemical evolution models \cite{ashenfelter04,fenner05}.

Other sources of systematics in the MM method include the assumption 
that all species have the same kinematic structure and the possibilities of 
line blends and line misidentifications (e.g.  \cite{bahcall04,murphy01d,murphy03}). 
As emphasized by \cite{murphy01d}, these are likely to result in random contributions 
to $\dal$. However, it is unclear whether the number of systems is
sufficient to cancel such systematic effects at the level of $(v/c) \sim 0.1$, i.e. 
$\Delta v \sim 0.3$~km/s \cite{bahcall04}. Note that the SIDAM method does not
require the assumption that different species have the same velocity structure 
\cite{molaro07}, an advantage over the original MM method. Finally, the fact that the 
transitions used in the MM and SIDAM methods lie at rest-frame ultraviolet wavelengths 
has meant that it is difficult to apply these to Galactic lines of sight, to 
test whether the expected null result is obtained at $z = 0$.

\subsection{Molecular hydrogen lines}
\label{sec:htwo}

The numerous ultraviolet ro-vibrational transitions of molecular hydrogen (\htwo)
have different dependences on the reduced molecular mass, implying that comparisons
between the line redshifts can be used to probe changes in $\mu$ \cite{thompson75,varshalovich93}.
Unfortunately, these lines are both weak and, for redshifted absorbers, lie in the 
Lyman-$\alpha$ forest, making them difficult to detect. As a result, only about a dozen 
redshifted \htwo\ absorbers are presently known, of which only three have been useful 
in studying changes in $\mu$ (e.g. \cite{varshalovich93,reinhold06,king08}).
While all earlier studies had provided constraints on $\dmu$, \cite{reinhold06} used 
VLT-UVES observations of two of these systems to obtain $\dmu = (2.0 \pm 0.6) \times 10^{-5}$ 
over $0 < z < 3$, i.e. $3.5\sigma$ evidence for a larger value of $\mu$ at early times. 
A recent re-analysis of the same data, combined with VLT-UVES data on the third 
absorber, at $z \sim 2.811$ towards PKS~0528$-$25, has yielded the strong constraint 
$\dmu < 6.0 \times 10^{-6}$ \cite{king08}.  However, the new (high-sensitivity) data 
towards PKS~0528$-$25 show a complicated velocity structure and the result could be affected 
by the presence of unresolved spectral components. In addition, \cite{king08} choose to carry 
out a simultaneous fit to the \htwo\ profiles as well as all Lyman-$\alpha$ forest transitions 
near the \htwo\ lines, rather than excluding possible blends with Lyman-$\alpha$ forest 
interlopers (as is usually done; e.g. \cite{reinhold06}. While this allowed them to retain 
a far larger number of \htwo\ lines in the analysis, it is unclear what effect it might have 
for their results. Again, it has not been possible to test that this method yields the expected 
null result in the Galaxy.

\section{Radio techniques}
\label{sec:radio}

The methods discussed in the preceding section were all based on optical spectroscopy,
where relative wavelength calibration of echelle orders, line blending due to the low 
spectral resolution ($\sim 6-8$~km/s), local velocity offsets, line interlopers, etc, are all 
possible sources of systematic error (e.g. \cite{murphy03}). The many-multiplet method
is affected by the additional problem of unknown isotopic abundances at high redshift, 
while fits to redshifted \htwo\ transitions must be carried out in the fog of 
the Lyman-$\alpha$ forest. Given the possibility that under-estimated or unknown 
systematics might dominate the errors from a given technique, it is important that 
independent techniques, with entirely different systematic effects, be used. 

Different radio transitions arise from very different physical mechanisms (e.g. hyperfine 
splitting, molecular rotation, Lambda-doubling, etc) and the line frequencies hence have 
different dependences on the fundamental constants. Comparisons between the redshifts of 
radio lines (or between radio and optical lines; \cite{wolfe76}) can thus be used 
to probe changes in the fundamental constants, with the immediate benefit that systematic
effects are very different in the radio regime. Obvious advantages over optical techniques 
are the high spectral resolution ($< 1$~km/s) possible at radio frequencies, alleviating 
problems with line blending, and the fact that the frequency scale is set by accurate 
masers and local oscillators, allowing frequency calibration to better than $\sim 10$~m/s. 
Further, comparisons between the redshifts of multiple transitions from a single molecule 
(e.g. OH, NH$_3$, CH), with different dependences on the constants, reduces the likelihood
that the result might be affected by local velocity offsets in the gas cloud between the lines.

\subsection{The HI~21cm hyperfine transition}
\label{sec:hi-opt}

The \hi~21cm hyperfine transition is the most commonly detected radio line 
at cosmological distances and is hence used in a number of comparisons to 
probe changes in the fundamental constants. \cite{wolfe76} were the first 
to use the \hi~21cm line for this purpose, comparing \hi~21cm and metal-line 
(Mg{\sc ii}) absorption redshifts in a $z \sim 0.524$ absorber to constrain 
evolution in the quantity $X \equiv g_p \mu \alpha^2$. More recently, 
\cite{tzanavaris07} used a similar comparison to obtain $[\Delta X/X] < 2 \times 10^{-5}$ 
from nine redshifted \hi~21cm absorbers at $0.23 < z < 2.35$. 
An obvious concern about this measurement is that \cite{tzanavaris07} compared 
the redshifts of the deepest \hi~21cm and metal-line absorption components, 
assuming that these arise in the same gas. However, many of the systems used 
in \cite{tzanavaris07} have multiple components in both \hi~21cm and metal 
transitions and it is by no means necessary that the strongest absorption 
arises in the same component in both types of lines \cite{kanekar06}.

Comparisons between the redshifts of the \hi~21cm and molecular rotational transitions 
(e.g. CO, HCO$^+$) are sensitive to changes in $Y \equiv g_p \alpha^2$. At present, there are 
just four redshifted systems with detections of both \hi~21cm and rotational 
molecular absorption (e.g. \cite{wiklind97,carilli97}), of which 
only two are suitable to probe changes in the constants. \cite{carilli00} used these to obtain 
$[\Delta Y/Y] < 3.4 \times 10^{-5}$ from $z \sim 0.25$ and $z \sim 0.685$, conservatively 
assuming that differences between the \hi~21cm and millimetre sightlines could 
yield local velocity offsets of $\sim 10$~km/s.

An important source of systematic effects in both these techniques is that the 
background quasar is far larger at the low redshifted \hi~21cm frequencies 
($\lesssim 1$~GHz) than at the millimetre wavelengths of the strong rotational 
transitions or the optical fine structure lines. This implies that the sightlines 
in the different transitions could probe different velocity structures in the 
absorbing gas. For example, large velocity offsets ($\sim 15$~km/s) have been 
observed between the \hi~21cm and HCO$^+$ redshifts at $z \sim 0.674$ towards 
B1504+377, probably due to small-scale structure in the absorbing gas 
\cite{wiklind96,carilli97,kanekar08c}. This is less of an issue for a comparison 
between the \hi~21cm and ``main'' OH~18cm lines, which is sensitive to changes in 
$Z \equiv g_p[\mu\alpha^2]^{1.57}$ \cite{chengalur03}. \cite{kanekar05} used four 
\hi~21cm and OH~18cm spectral components in two gravitational lenses to obtain 
$[\Delta Z/Z] < 2.1 \times 10^{-5}$ from $z \sim 0.7$, assuming possible systematic
velocity errors of $\sim 3$~km/s between the \hi~21cm and OH components. The strong 
dependence of $Z$ on $\alpha$ ($Z \propto \alpha^{3.14}$) implies that this 
comparison is very sensitive to changes in $\alpha$, with  a $2\sigma$ sensitivity of
$\dal < 6.7 \times 10^{-6}$ \cite{kanekar05}.

\subsection{Ammonia inversion transitions}
\label{sec:nh3}

An interesting technique to probe changes in the proton-electron mass ratio, 
using inversion transitions of the ammonia (NH$_3$) molecule, was proposed by 
\cite{flambaum07b}. The dependence of the NH$_3$ inversion frequency on $\mu$ is 
far stronger than that of rotational line frequencies, implying that a comparison 
between NH$_3$ and (for example) CO redshifts in an absorber can yield a high 
sensitivity to changes in $\mu$. Recently, \cite{murphy08} used NH$_3$ and 
rotational (HCO$^+$ and HCN) lines in the $z \sim 0.685$ gravitational lens 
towards B0218+357 to obtain $\dmu < 1.8 \times 10^{-6}$ over $\sim 6.5$~Gyr. 
While this is the most sensitive single bound on changes in $\mu$, the analysis 
assumes the same velocity structure in the NH$_3$ and rotational 
lines (8 spectral components), despite the fact that there is no evidence for more 
than 2 components in the NH$_3$ spectra \cite{henkel05}. It is unclear 
what effect this might have on the results. Further, the NH$_3$ lines are assumed to be 
in local thermodynamic equilibrium, so that the eighteen components of the NH$_3$~1-1 
multiplet have the same velocity width; it is not possible to verify this at the low 
signal-to-noise ratio (S/N) of the present NH$_3$ spectrum. The quasar image (``A''; 
\cite{wiklind99a}) that shows strong molecular absorption is also only partly obscured at 
optical wavelengths (remaining clearly visible), despite the high extinction of the 
foreground molecular cloud.  This indicates structure in the molecular cloud on the very 
small scales of the optical quasar \cite{wiklind99a}, which could lead to different velocity 
structures in the NH$_3$ and HCO$^+$ lines, due to their very different line 
frequencies ($\sim 23.7$~GHz and $\sim 178.4$~GHz, in the absorber's rest frame; \cite{murphy08}).  
In addition, the analysis assumes no local velocity offsets between the NH$_3$ and rotational 
lines. As always, comparisons between lines from different species require large absorber 
samples to average out such effects.

\section{``Conjugate'' Satellite OH lines}
\label{sec:oh}

The satellite OH~18cm lines are said to be conjugate when the two lines have the 
same shape, but with one line in emission and the other in absorption. 
This arises due to an inversion of the level populations within the ground state of 
the OH molecule \cite{elitzur92,langevelde95}. For clarity, Fig.~\ref{fig:ohlevels} 
shows the three lowest OH rotational states (not to scale). The $^2\Pi_{3/2}$ J=3/2 OH 
ground rotational state (and every excited OH state) is split up into four sub-levels 
by $\Lambda$-doubling and hyperfine splitting; transitions between these sub-levels give 
rise to the microwave OH spectrum.  Transitions with $\Delta F = 0$, i.e. with no change in 
the total angular momentum quantum number $F$, are referred to as ``main'' lines, while 
transitions with $\Delta F = \pm 1$ are 
called ``satellite'' lines. The rest frequencies of the ground state lines are known to 
high accuracy, with the main lines at 1665.401803(12)~MHz and 1667.358996(4)~MHz 
\cite{hudson06}, and the satellite lines at 1612.230825(15)~MHz and 1720.529887(10)~MHz 
\cite{lev06}. 

\setcounter{figure}{0}
\begin{figure}[t!]
\begin{center}
\includegraphics[height=5.0in,angle=270]{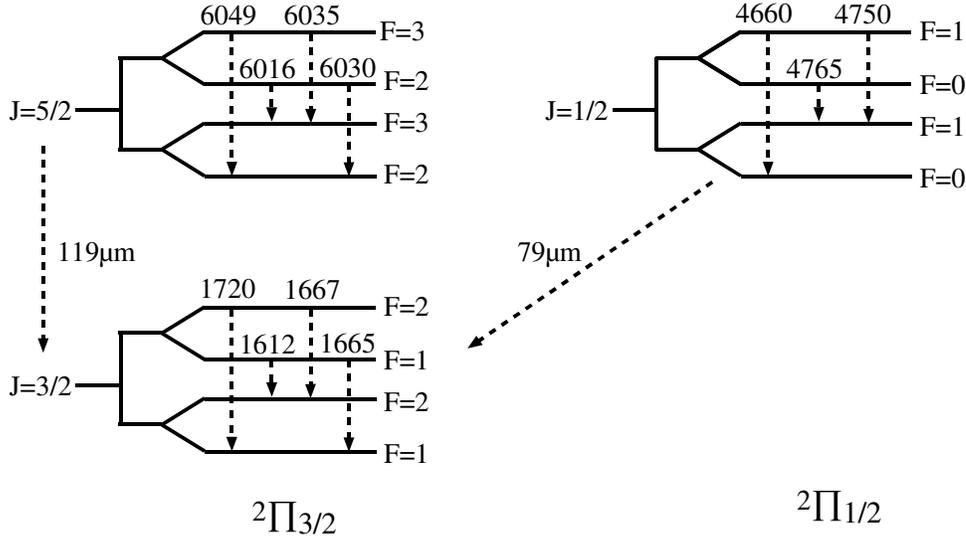}
\end{center}
\caption{The 3 lowest rotational OH states (not to scale) and their microwave and infra-red lines.}
\label{fig:ohlevels}
\end{figure}

If the OH molecules are pumped from the ground state to higher rotational states, either
collisionally or radiatively, they cascade back down to the ground state. The last 
step of the cascade can be either the 119$\mu$m intra-ladder transition, 
{$2\Pi_{3/2}$~(J=5/2)}~$\rightarrow$~{$2\Pi_{3/2}$~(J=3/2)} or the 79$\mu$m cross-ladder 
transition, {$2\Pi_{1/2}$~(J=1/2)}~$\rightarrow$~~{$2\Pi_{3/2}$~(J=3/2)}. Both decays are 
governed by the quantum mechanical selection rules $\Delta F = 0,\: \pm 1$. For
the $119\mu$m route, this means that transitions from the $F=3$ sub-levels of the excited state 
to the $F=1$ sub-levels of the ground state are forbidden, while the $F=2$ sub-levels 
can decay to all ground state sub-levels. Conversely, for the $79\mu$m route, transitions 
between the $F=1$ excited state sub-levels and the $F=3$ ground state sub-levels are forbidden, 
while all other transitions are allowed. Further, if the $119\mu$m or $79\mu$m transitions 
are optically thick (requiring $N_{\rm OH} \gtrsim 10^{15}$~cm$^{-2}$; \cite{elitzur92}), 
the sub-level populations in the ground state are determined only by 
the number of possible routes to each sub-level \cite{elitzur92}. Thus, a cascade ending
with an optically-thick $119\mu$m decay will result in an over-population of the $F=2$ ground
state sub-levels, while one ending in an optically-thick $79\mu$m decay will yield an 
over-population of the $F=1$ sub-levels. Neither of these would have any effect on the ground state 
``main'' lines, as the latter connect sub-levels with the same total angular momentum. 
However, the situation is very different for the satellite lines, connecting states with 
different $F$ values. The first case (an optically-thick $119\mu$m decay), would result in 
stimulated 1720~MHz emission and 1612~MHz absorption, while the second case (an optically-thick 
$79\mu$m decay) would give rise to 1612~MHz emission and 1720~MHz absorption. In both cases, the 
satellite lines would also have the same shapes, implying that the sum of the satellite optical 
depths would be consistent with noise \cite{elitzur92,langevelde95}.

Such ``conjugate'' satellite behaviour has been seen in nearby extra-galactic systems
for more than a decade (e.g. Centaurus~A; \cite{langevelde95}) but has only recently been 
discovered at cosmological distances \cite{kanekar04b,darling04,kanekar05}. This is extremely
useful for studies of fundamental constant evolution because the four OH~18cm transition 
frequencies have very different dependences on $\alpha$, $\mu$ and $g_p$ \cite{chengalur03}. 
Specifically, a comparison between the sum and difference of satellite line redshifts probes 
changes in the quantity $G \equiv g_p \left[ \mu \alpha^2 \right]^{1.85}$ \cite{chengalur03}, with
a high sensitivity to changes in both $\alpha$ and $\mu$ (a similar analysis by \cite{darling03} 
implicitly assumes no changes in $\mu$ and $g_p$). Note that \cite{kanekar04b,chengalur03}
define $\mu \equiv m_e/m_p$; the more common $\mu \equiv m_p/m_e$ is used here.

Only two redshifted conjugate satellite OH systems are currently known, at $z \sim 0.247$ 
towards PKS~1413+135 \cite{kanekar04b,darling04} and $z \sim 0.765$ 
towards PMN~J0134$-$0931 \cite{kanekar05}. We have recently carried out deep observations of 
the satellite OH lines of PKS~1413+135 with the Westerbork Synthesis Radio Telescope (WSRT) and 
the Arecibo telescope \cite{kanekar08b}. The WSRT and Arecibo optical depth spectra have root-mean-square 
noise values of $\sim 8.5 \times 10^{-4}$ per $\sim 0.55$~km/s and $\sim 7 \times 10^{-4}$ per $\sim 0.36$~km/s, 
respectively, after Hanning smoothing and re-sampling. These correspond to S/N$\sim 1200$ at a  
resolution $R \sim 550000$ and S/N$\sim 1430$ at $R \sim 830000$; for comparison, 
the 143~Keck-HIRES spectra of \cite{murphy04} typically had S/N$\sim 30$ at $R \sim 45000$. 
The sum of the 1612 and 1720~MHz optical depth profiles was found to be consistent 
with noise in both data sets, as expected for conjugate behaviour, with no evidence
for a velocity offset between the satellite lines. Our preliminary result from a 
cross-correlation analysis of the two data sets is $\left[ \Delta G/G\right] < 1.1 \times 10^{-5}$, 
where $G \equiv g_p\left[\mu \alpha^2\right]^{1.85}$, consistent with no changes in 
$\alpha$, $\mu$ and $g_p$ from $z \sim 0.247$ to the present epoch. Assuming that changes 
in $g_p$ are much smaller than those in $\alpha$ or $\mu$ (e.g. \cite{langacker02}), this
gives $2\sigma$ sensitivities of $\dal < 3.1 \times 10^{-6}$ or $\dmu < 6.2 \times 10^{-6}$,
over $0 <  z < 0.247$. The data are now being inspected for possible systematic 
effects.

We have also applied this technique to OH~18cm data from a nearby conjugate satellite 
system, Centaurus~A \cite{langevelde95}.
Here, the cross-correlation of the two spectra peaks at $\Delta V = (-0.05 \pm 0.11)$~km/s, 
consistent with no velocity offset between the lines. This yields $\left[\Delta G/G\right] <
1.16 \times 10^{-5}$, from $z \sim 0.0018$. This demonstrates the expected null result 
in a ``local'' comparison, at a sensitivity similar to that of the high-$z$ result.

The result for PKS~1413+135 can be used to obtain (model-dependent) constraints on 
both $\dal$ and $\dmu$ by defining 
$\left[ \Delta \mu/\mu \right] = R \left[ \Delta \alpha/ \alpha \right]$. This yields
$\left[ \Delta \alpha/ \alpha \right] \times (2 - R) = (+5.0 \pm 3.1) \times 10^{-6}$, 
again assuming $\left[\Delta g_p/g_p\right] << \dal, \dmu$. Given a value of $R$ from 
a theoretical model, one can immediately obtain an estimate for $\left[ \Delta \alpha/\alpha \right]$. 
For example, values of $R \sim -50$ are typical in GUT models (e.g.  \cite{langacker02,calmet02}), 
giving $\left[ \Delta \alpha/ \alpha \right] \lesssim few \times 10^{-7}$ (note that $R$ can 
have either sign but is typically large in absolute value, in the absence of fine-tuning). 
Changes in $G$ are then dominated by changes in $\mu$; any model with $|R| \gtrsim 10$ yields 
a similar result.  This gives $\left[ \Delta \alpha/ \alpha \right] < 8 \times 10^{-7}$ and 
$ \left[ \Delta \mu/\mu \right] \lesssim 8 \times 10^{-6}$ from $z \sim 0.247$.

The above discussion also highlights the main drawback of the conjugate-satellites technique,
the fact that it cannot be used to directly measure changes in individual constants 
(like the many-multiplet or \htwo\ methods), but instead probes changes in a combination 
of three constants, $\alpha$, $\mu$ and $g_p$. Constraints on changes in the individual 
constants from this method are hence model-dependent (e.g. the assumption that $g_p$ 
remains unchanged). \cite{kanekar04b} point out that, for the $119\mu$m decays,
an over-population of the ground state $F=2$ sub-levels should be accompanied by an 
over-population of the $F=3$ sub-levels in the $J=5/2$ state, implying that the $J=5/2$ satellite
lines should also be conjugate. Using these lines in tandem with the ground state satellite
lines would reduce the degeneracy in the conjugate-satellites method between changes in 
$\alpha$, $\mu$ and $g_p$ \cite{kanekar04a}. However, despite deep searches \cite{kanekar08b}, 
the $J=5/2$ satellite lines have not so far been detected at cosmological distances.

The strength of the conjugate-satellites technique stems from the fact that the conjugate 
behaviour {\it guarantees that the satellite lines arise from the same gas}. Such systems 
are hence perfectly suited to probe changes in $\alpha$, $\mu$ and $g_p$ from the source 
redshift to today, as systematic velocity offsets between the lines are ruled out by the inversion 
mechanism. Any measured difference between the line redshifts must then arise due to a 
change in one or more of $\alpha$, $\mu$ and $g_p$ \cite{kanekar04b}. The technique 
also contains a stringent test of its own applicability, in that the shapes of the two 
lines must agree if they arise in the same gas. Further, it provides a estimate of changes 
in the fundamental constants from a single space-time location, without any averaging over 
multiple absorbers (which, for example, is essential in most other methods to average out 
local systematics). In addition, the velocity offset between the lines can be determined from
a cross-correlation analysis, as the shapes of the lines are the same; it is not necessary 
to model the line profiles with multiple spectral components of an assumed shape, 
which can, especially for complex profiles, itself affect the result. As noted earlier, 
the use of masers of local oscillators to set the radio frequency scale means that 
frequency calibration is also not an issue here. Finally, the nearest isotopic OH transitions 
are more than $50$~MHz away and this part of the radio spectrum has very few other known 
astronomical transitions; line interlopers are thus not likely to be an issue in this 
method, unlike the situation in the optical regime \cite{kanekar08b}. Overall, systematic 
effects appear to be far less important for the conjugate-satellites method than for 
the other techniques. 

\section{Results from the different techniques}
\label{sec:present}

\setcounter{figure}{1}
\begin{figure*}[t!]
\epsfig{file=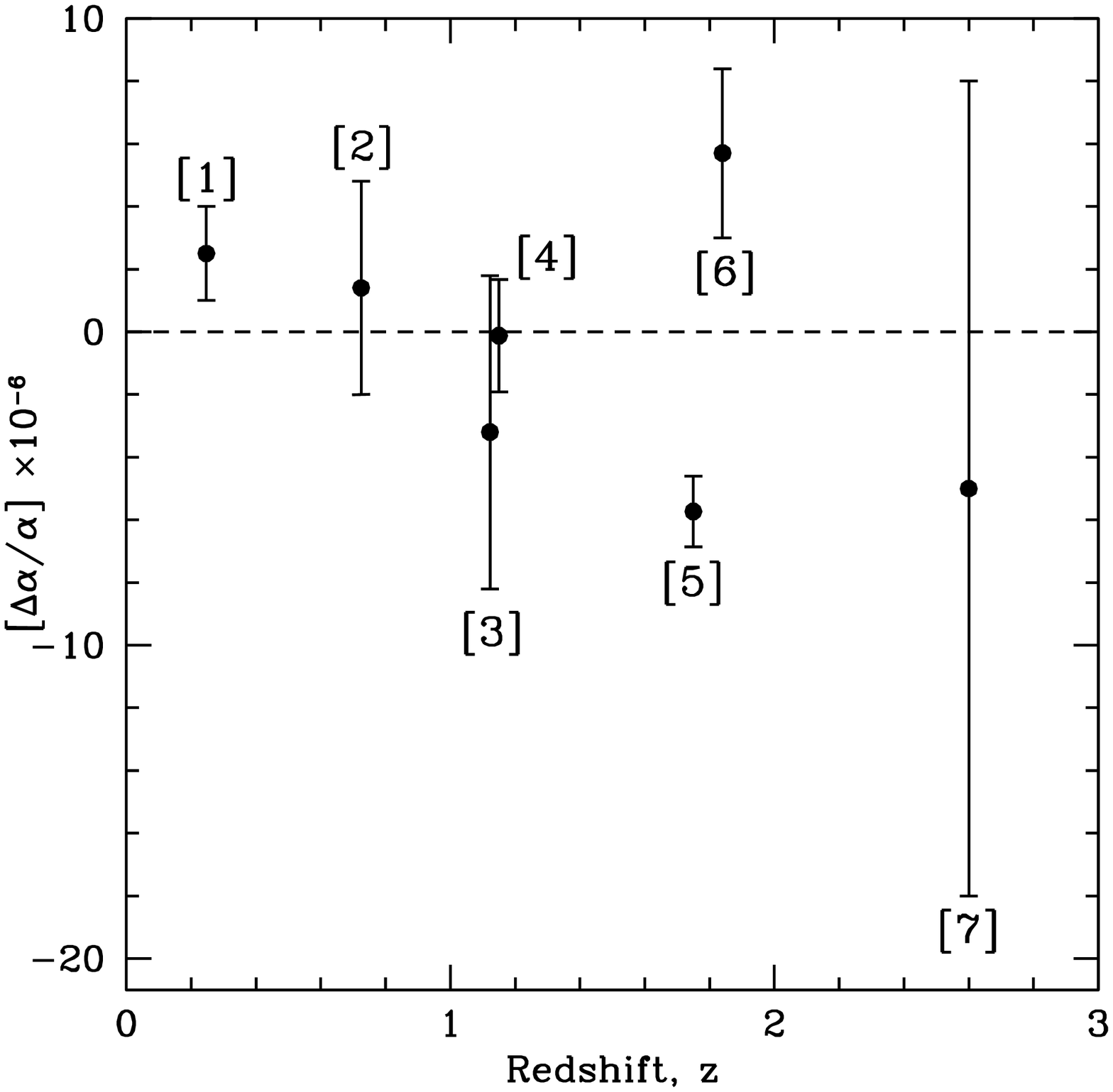,width=2.5in}
\vskip -2.5in
\hskip 2.5in
\epsfig{file=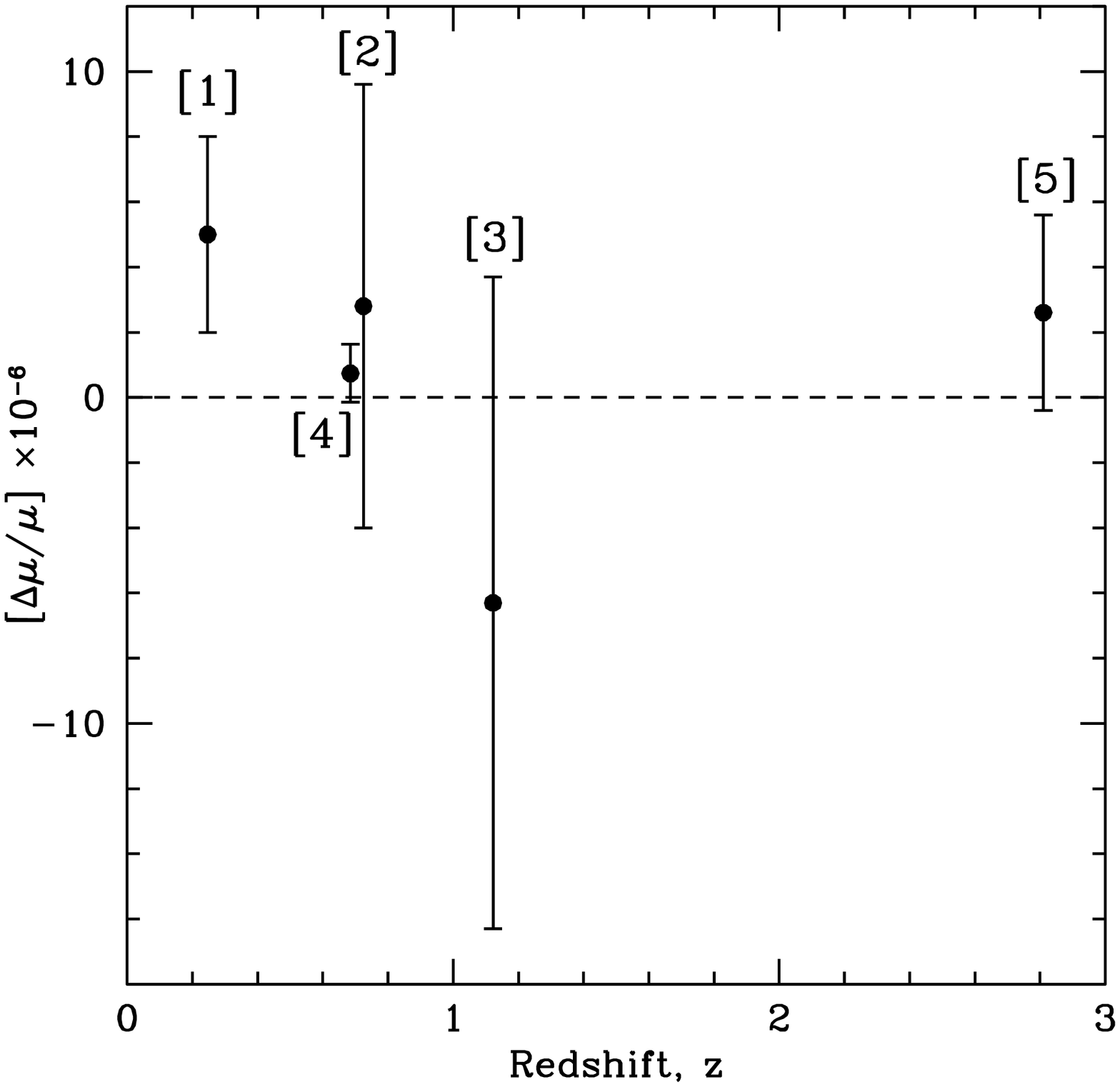,width=2.5in}
\caption{Current results from techniques probing fundamental constant evolution 
with redshifted spectral lines. The left panel [A] shows the best present 
estimates for $\dal$ as a function of redshift, from the [1]~conjugate-satellites \cite{kanekar08b}, 
[2]~\hi~21cm vs. OH lines \cite{kanekar05}, [3]~\hi~21cm vs. fine structure lines \cite{tzanavaris07}, 
[4]~SIDAM, at $z \sim 1.15$ \cite{molaro07}, [5]~MM \cite{murphy04}, [6]~SIDAM, at $z \sim 1.84$ 
\cite{molaro07}, and [7]~AD \cite{murphy01c}, methods. Panel~[B] shows similar results 
for $\dmu$, from the [1]~conjugate-satellites \cite{kanekar08b}, [2]~\hi~21cm vs. OH lines 
\cite{kanekar05}, [3]~\hi~21cm vs. fine structure lines \cite{tzanavaris07}, 
[4]~NH$_3$ vs. rotational lines \cite{murphy08}, and [5]~\htwo \cite{king08}, methods.
The assumptions $\dal >> \dmu, \left[\Delta g_p/g_p\right]$ (in [A]) and $\left[\Delta g_p/g_p\right], 
\dal << \dmu$ (in [B]) apply to the first three in each 
category. Note that the comparison between NH$_3$ and rotational lines assumes no local 
velocity offsets between the lines, while the \hi~21cm vs. OH comparison uses a 
velocity dispersion of 3~km/s \cite{kanekar05,murphy08}.}
\label{fig:full}
\end{figure*}

It is difficult to directly compare results from radio and optical techniques as the 
former (except for the NH$_3$ lines) probe combinations of $\alpha$, $\mu$ and $g_p$.
A model-dependent summary of the best present results can be obtained for the two limiting cases, 
$\dal >> \dmu$ and $\dal << \dmu$, assuming that $\left[ \Delta g_p/g_p\right] << \dal, \dmu$ 
(e.g. \cite{langacker02}). Such a comparison, between the alkali doublet, many-multiplet, SIDAM, 
\hi~21cm vs. fine structure, \hi~21cm vs. OH, NH$_3$ vs. rotational, conjugate-satellites and 
\htwo\ methods, is shown in the two panels of Fig.~\ref{fig:full} 
\cite{murphy01c,murphy03,king08,tzanavaris07,kanekar05,kanekar08b,murphy08}.
It is apparent that present results from the conjugate-satellites, many-multiplet and SIDAM
methods have similar sensitivities to changes in $\alpha$, although at very different 
lookback times. Conversely, the NH$_3$ method is the most sensitive among probes of changes 
in $\mu$ (although note that \cite{murphy08} assume no local offsets between the NH$_3$ 
and rotational lines), with the conjugate-satellites and \htwo\ methods having 
similar sensitivities, albeit again at widely separated redshifts.

The figure also shows a clear separation between the redshift ranges at which 
the radio and optical techniques have been applied. The optical techniques are all based on 
ultraviolet transitions that only move into optical wavebands (say, $\gtrsim 3200$\AA) for
absorbers beyond a certain redshift. This means that, for example, the H$_2$ technique can 
only be used at $z \gtrsim 2$, the Si{\sc iv} AD technique at $z \gtrsim 1.3$ and the SIDAM 
method (using the Fe{\sc ii}$\lambda$1608 line; \cite{quast04}) at $z \gtrsim 1$. Similarly, only a handful 
of lines from singly-ionized species are redshifted above 3200\AA\ at $z \lesssim 0.6$, 
implying that the many-multiplet method too works best at higher redshifts ($z \gtrsim 0.8$). 
On the other hand, the lack of known radio molecular absorbers at $z > 0.9$ 
limits the radio methods to relatively low redshifts. Present radio and optical techniques thus play 
complementary roles in studies of fundamental constant evolution, with the best low-$z$ 
measurements coming from radio wavebands and the best high-$z$ estimates, from the optical 
regime. It is also clear that the many-multiplet method is the only present technique 
that finds statistically significant evidence of changes in $\alpha$ or $\mu$.

\section{Future studies}
\label{sec:future}

The present limitations of the optical and radio techniques are very different in nature.
The optical methods are affected by issues like line blending, relative calibration of 
different echelle orders, unknown relative isotopic abundances at high redshifts and 
the fact that it has not been possible to obtain an estimate of local systematic 
effects in the Galaxy. The biggest drawback to the radio methods is the paucity 
of high-$z$ radio absorbers, due to which one cannot average out local systematics 
(note that this does not affect the conjugate-satellites method).

Over the next decade, new telescopes and associated instrumentation will address some 
of these issues, at both radio and optical wavebands. The next generation of (30-metre-class) 
optical telescopes and new spectrographs should result in both improved spectral resolution 
(alleviating problems with line blends, especially useful for the \htwo\ method) as well 
as much better wavelength calibration. However, even the planned spectrograph CODEX, with a 
resolving power of $R \sim 150000$ (e.g. \cite{molaro06}), will be unable to resolve out the 
Mg isotopic structure (the different isotopic transitions are separated by only $\sim 0.85$~km/s 
and $\sim 0.4$~km/s for Mg{\sc ii}$\lambda$2803 and Mg{\sc I}$\lambda$2853, respectively; 
\cite{murphy01c}). Unknown relative isotopic abundances in the high-$z$ absorbers 
is thus likely to remain an issue for the MM (and SIDAM) methods, although statistical errors
of $\dal \sim 10^{-7}$ should be achievable. 
Additional corrections to the Born-Oppenheimer approximation may be necessary to 
improve the precision in the \htwo\ method \cite{reinhold06} to reach similar sensitivities 
in $\dmu$. Finally, it is unlikely that it will be possible to test for a null result from 
these methods in the Galaxy, as this would require high-resolution ultraviolet spectroscopy.

On the other hand, increasing the number of redshifted systems detected in atomic and 
molecular radio lines is a critical ingredient for future radio studies of fundamental 
constant evolution. 
The wide-band receivers and correlators of the Green Bank Telescope, the Atacama Large 
Millimeter Array and the Expanded Very Large Array will, for the first time, allow ``blind'' 
surveys for redshifted absorption in the strong mm-wave rotational transitions towards a large 
number of background sources. This should yield sizeable samples of high-$z$
absorbers in these transitions, which can then be followed up in other molecular 
and atomic lines. It should then be possible to average over local velocity offsets 
or different kinematic structures to obtain reliable results when comparing line redshifts 
across species. These surveys should also reveal a new population of redshifted 
conjugate satellite OH systems (note that two of the five known radio molecular 
absorbers show conjugate-satellite behaviour). Deep ($ > 100$~hour) integrations on the 
$z \sim 0.247$ conjugate system in PKS~1413+135 with existing telescopes in the next few years 
should achieve $1\sigma$ sensitivities of $\dal, \dmu \sim few \times 10^{-7}$ while, in the 
long-term future, the Square Kilometer Array will be able to detect fractional changes of 
$\dal \sim 10^{-7}$ in both known (and any newly-detected) conjugate systems.

It thus appears that similar sensitivity to fractional changes in $\dal$ will be 
achievable with both the many-multiplet (or SIDAM) and conjugate-satellites methods 
over the next decade or so, although new conjugate systems will have to be detected at 
high redshifts to extend this method beyond $z \sim 0.8$. The lack of apparent 
systematic effects in the conjugate-satellites method suggest that it is likely to
be the most reliable probe of fundamental constant evolution, unless the issue of 
unknown isotopic abundances in the MM method can be resolved. However, the drawback 
that the conjugate-satellites method is only sensitive to changes in a combination of 
$\alpha$, $\mu$ and $g_p$ means that other approaches will be necessary to measure 
changes in these constants independently. In any event, it remains important that multiple 
techniques be used so as to ensure that the results are not affected by unknown systematic 
effects that could be specific to one method.

\section{Summary}
\label{sec:summary}

Optical and radio techniques have played complementary roles in extending studies 
of fundamental constant evolution to large lookback times. Optical approaches
have provided high sensitivity at high redshifts, $z \sim 0.8-3.0$, while radio 
approaches have been critical to study the late-time ($z \lesssim 0.8$) behaviour.
The best present results, from a variety of methods, have $2\sigma$ sensitivities 
of $\dal, \: \dmu \sim few \times 10^{-6}$ at $z \sim 0.25 - 2.8$, an improvement 
of nearly two orders of magnitude in the last decade. The conjugate-satellites
technique appears to have the fewest systematic effects of all methods considered here,
with the agreement between the satellite OH line shapes providing a stringent test 
of its own applicability. The only technique yielding statistically-significant evidence for 
changes in the fundamental constants remains the many-multiplet method \cite{murphy04}, which
finds a smaller value of $\alpha$ at early times. While it has still not been possible to 
definitely confirm or deny this result, much progress has been made in developing new 
techniques and understanding systematic effects. It is important to continue such studies 
with a range of techniques, both to ensure that the results are not dominated 
by systematic effects in any given approach and to probe changes in multiple constants
over a wide range of redshifts.

\section{Acknowledgments}
\label{sec:ack}
It is a pleasure to thank Jayaram Chengalur and Carlos Martins for stimulating 
discussions on fundamental constant evolution and comments on an earlier draft, 
Jayaram and Tapasi Ghosh for permission to discuss our unpublished results, and
Huib van Langevelde for providing us with the OH spectra towards Cen.A.
I acknowledge support from the Max Planck Foundation and an NRAO Jansky Fellowship.

\bibliographystyle{prsty}
\bibliography{ms}
\end{document}